\documentclass[aps,twocolumn,showpacs]{revtex4}%
\usepackage{amsfonts}
\usepackage{amsmath}
\usepackage{amssymb}
\usepackage{appendix}
\usepackage{graphicx}%
\providecommand{\U}[1]{\protect\rule{.1in}{.1in}}
\providecommand{\U}[1]{\protect\rule{.1in}{.1in}}

\begin{document}
\title{Optomechanically induced optical responses with non-rotating wave approximation}
\author{\ Xiao-Bo Yan}
\email{xiaoboyan@126.com}
\affiliation{College of Electronic Science, Northeast Petroleum University, Daqing  163318, China}
\date{\today}


\begin{abstract}
Slow light propagation is an important phenomenon in quantum optics. Here, we theoretically study the properties of slow light in a simple optomechanical system considering an effect of non-rotating wave approximation (NRWA) that was ignored in previous related works. With the NRWA effect, the ultraslow light can be easily achieved at the window of optomechanically induced transparency, especially in unresolved sideband regime. From the theoretical results, we find the upper bound of the time delay is exactly the mechanical ringdown time which can last for \textit{several minutes} (mHz linewidth) in recent experiments. Additionally, the interesting phenomena of the perfect optomechanically induced transmission and absorption are studied in the system with the NRWA effect. We believe the results can be used to control optical transmission in modern optical networks.
\end{abstract}
\maketitle

\section{Introduction}

Cavity optomechanics \cite{Aspelmeyer2014} exploring the interaction between
macroscopic mechanical resonators and light fields, has received increasing
attention for the broad applications in testing macroscopic quantum physics,
high-precision measurements, and quantum information processing
\cite{Aspelmeyer2014,Kippenberg2008,Marquardt2009,Verlot2010,Mahajan2013}.
Various experimental systems exhibiting such
interactions are proposed and investigated, such as Fabry-Perot cavities \cite{Gigan2006,Arcizet2006}, whispering-gallery microcavities \cite{Kippenberg2005,Tomes2009,Jiang2009}, superconducting circuits \cite{Regal2008,Teufel2011_471}, and membranes \cite{Thompson2008,Jayich2008,Sankey2010,Karuza2013}.
The motion of mechanical oscillator and the optical properties in these optomechanical systems can be strongly affected by the interaction, and then various interesting quantum phenomena can be generated, such as ground-state cooling of mechanical
modes
\cite{Marquardt2007,Wilson-Rae2007,BingHe2017}, nonclassical mechanical states \cite{Nation2013,Ren2013}, photon antibunching
\cite{Rabl2011prl,Xu2013pra,Wang2019pra,Liao2013pra}, and optomechanically induced transmission and absorption \cite{Yan2014,Qu2013pra,Agarwal2014njp,Zhang2017pra,Yan2019FOP}.

Recently, the study of optomechanically induced transparency (OMIT) has attracted much attentions \cite{Huang2010_041803,Weis2010,Yan2015,Safavi-Naeini2011,Huang2011,Jing2015,LiuYC2017,Yan2020pra,Shahidani2013,Chen2011,LiuYX2013,Kronwald2013,Lu2017,Lu2018,Dong2013,Dong2015,Ma2014pra,Xiong2012,Xiong2018}.
It was theoretically predicted by Agarwal and Huang \cite{Huang2010_041803} and experimentally observed in a microtoroid system \cite{Weis2010}, a membrane-in-the-middle system \cite{Karuza2013}, and in a nanoscale optomechanical crystal \cite{Safavi-Naeini2011}. 
A remarkable feature of optomechanically induced transparency is the drastic reduction in the group velocity of light passing through the system \cite{Safavi-Naeini2011}, achieved due to the abnormal dispersion accompanied with the transparency
window. This aspect of the effect has been utilized to conjure schemes whereby light may be slowed and stopped \cite{Chen2011,Chang2011njp,Tarhan2013,Akram2015,Gu2015,Safavi-Naeini2011}, making it an important building block in quantum information and communication proposals, as well as of great practical interest in classical optics and photonics. 

In general, the narrower the transparent window is, the more significant the abnormal dispersion is.
Hence, it is important for the abnormal dispersion to have both a large transparency depth and a narrow transparency window. However, the ideal depth of the transparency window cannot be achieved due to the nonzero mechanical damping rate in the usual theory of OMIT \cite{Huang2010_041803,Safavi-Naeini2011,Weis2010,LiuYC2017,Xiong2018}, where the width of the transparency window is
very large due to the large driving strength needed to increase the depth
of transparency. Hence, the slow light effect based on the usual OMIT theory must be quite limited.
Until very recently, in Ref. \cite{Yan2020pra} we provided a new mechanism (considering the NRWA effect) by which the ideal OMIT can easily be achieved with an ultra-narrow transparency window. It is conceivable that under the new mechanism, the slow light effect will be greatly enhanced. In addition, other optical properties of the system can also be dramatically modified by the NRWA effect, such as, the interesting phenomena of the optomechanically induced transmission and absorption.

Here, we theoretically study the ultraslow light, perfect optomechanically induced transmission and absorption with a simple “membrane-in-the-middle” configuration (see Fig. 1) \cite{Bhattacharya2008,Thompson2008,Sankey2010}, considering the NRWA effect which is ignored in previous related works. After considering the NRWA effect, we find it has a strong impact on the absorptive and dispersive behavior of the optomechanical system to the probe field. First, we obtain the analytic expression of the time delay, and find there is only slow light in the system because the time delay is always positive. Secondly, the ultraslow light can be easily achieved with small mechanical dissipation rate (the time delay can be improved to about \textit{one second} even with the usual Hz linewidth) at the narrow transparency window, especially in unresolved sideband regime. Thirdly, there exist an upper bound of the time delay that is exactly the ringdown time of the mechanical oscillator, and in recent optomechanical experiments \cite{Norte2016prl,Reinhardt2016prx,Ghadimi2018sci,Tsaturyan2017NatN}, the ringdown time can last for \textit{several minutes} (mHz linewidth). In addition, we study the interesting phenomena of the perfect optomechanically induced transmission and absorption with the NRWA effect, and give the conditions under which these phenomena can be achieved in the system.

\section{Model and equations}

\begin{figure}[ptb]
	\includegraphics[width=0.42\textwidth]{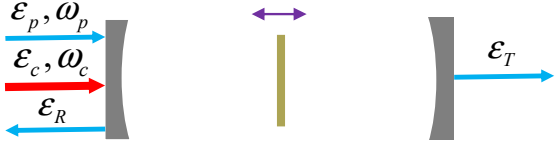}\caption{ Sketch of a membrane-in-the-middle optomechanical system consists of one mechanical membrane interacted with one cavity via radiation pressure effects. The cavity is driven by a coupling field with
		frequency $\omega_{c}$ (amplitude $\varepsilon_{c}$) and a weak probe field with frequency $\omega_{p}$ (amplitude $\varepsilon_{p}$).}%
	\label{Fig1}%
\end{figure}

We consider a membrane-in-the-middle optomechanical system in which a cavity with frequency $\omega_{0}$ and length $L$ is
coupled to a mechanical membrane with frequency $\omega_{m}$ and mass $m$ via radiation pressure effects (see Fig. 1).
The cavity annihilation (creation) operator is denoted by $c$ ($c^{\dagger}$) with the commutation relation $[c, c^{\dagger}]=1$, and the annihilation (creation) operator of the mechanical membrane is denoted by $b$ ($b^{\dagger}$) with $[b, b^{\dagger}]=1$.
If the mechanical membrane is placed at the node of the cavity field, the interaction Hamiltonian can be described by $-\hbar g_{0}c^{\dagger}c(b+b^{\dagger})$ and the optomechanical coupling rate $g_{0}=\frac{\sin (2kq_{0})}{\sqrt{2m\omega_{m}[(1-\mathcal{T})^{-1}-\cos^{2}(2kq_{0})]/\hbar}}(\frac{\omega_{0}}{L/2})$ \cite{Bhattacharya2008,Agarwal2014njp}, where $k$ is the wave vector of the cavity field, $q_{0}$ is the rest position of the membrane in the absence of radiation and $\mathcal{T}$ is the transmission rate of the membrane. The cavity is driven by a strong coupling field with
frequency $\omega_{c}$ (amplitude $\varepsilon_{c}$) and a weak probe field with frequency $\omega_{p}$ (amplitude $\varepsilon_{p}$). Then, the Hamiltonian in the rotating frame at the frequency of the coupling field $\omega_{c}$ is
\begin{eqnarray}
H&=&\hbar\Delta_{c}c^{\dagger }c+\hbar\omega_{m}b^{\dagger}b-\hbar g_{0}c^{\dagger}c(b^{\dagger}+b)\notag\\
&+&i\hbar
\varepsilon_{c}(c^{\dagger}-c)
+i\hbar(c^{\dagger}\varepsilon_{p}e^{-i\delta t}-c\varepsilon_{p}^{\ast}e^{i\delta t}).  
\end{eqnarray}%
Here, $\delta=\omega_{p}-\omega_{c}$ ($\Delta_{c}=\omega_{0}-\omega_{c}$) is the detuning between probe field (cavity field) and coupling field.

In this paper, we deal with the mean response of the system to the probe field in the presence of the coupling field, hence we do not include quantum fluctuations. 
We use the factorization assumption $\langle bc\rangle=\langle b\rangle\langle
c\rangle$ and then the mean value equations are then given by
\begin{eqnarray}
\langle\dot{b}\rangle &=&-\frac{\gamma}{2}\langle b\rangle-i\omega_{m}\langle
b\rangle+ig_{0}\langle c^{\dagger }\rangle \langle c\rangle,\\
\langle\dot{c}\rangle &=&-[2\kappa +i(\Delta_{c}-g_{0}\langle b^{\dagger}\rangle-g_{0}\langle b\rangle)]\langle c\rangle
+\varepsilon_{c}+\varepsilon_{p}e^{-i\delta t}.\notag
\end{eqnarray}
Here, $\gamma$ is the mechanical damping rate and $2\kappa$ is the cavity photon decay rate due to transmission losses through each end mirror of the cavity. 
In the absence of the probe field $\varepsilon_{p}$, the mean values of $\langle b\rangle$, $\langle c\rangle$ in the steady state can be obtained respectively as $b_{s}=ig_{0}|c_{s}|^{2}/(\frac{\gamma}{2}+i\omega_{m})$, $c_{s}=\varepsilon_{c}/(2\kappa+i\Delta)$ with $\Delta=\omega_{0}-\omega_{c}-g_{0}(b_{s}+b_{s}^{\ast})$.

In the presence of the probe field $\varepsilon_{p}$, we can write $\langle b\rangle=b_{s}+\delta b$ and $\langle c\rangle=c_{s}+\delta c$ to solve Eq. (2).
Substituting them into Eq. (2) and keeping only the
linear terms, we obtain the linearized Langevin equations
\begin{eqnarray}
\delta\dot{b}&=&-(\frac{\gamma}{2}+i\omega_{m})\delta b+ig_{0}(c_{s}^{\ast}\delta c+c_{s}\delta c^{\ast}),\\
\delta\dot{c}&=&-(2\kappa +i\Delta)\delta c+ig_{0}c_{s}(\delta b+\delta b^{\ast})+\varepsilon_{p}e^{-i\delta t}.\notag
\end{eqnarray}
Using the usual method \cite{Huang2010_041803,Weis2010,Safavi-Naeini2011,Karuza2013}, we can solve Eq. (3) by writing the solution in the form
$\delta b=\delta b_{+}e^{-i\delta t}+\delta b_{-}e^{i\delta t}$ and $\delta c=\delta c_{+}e^{-i\delta t}+\delta c_{-}e^{i\delta t}$. Inserting them into Eq. (3) and comparing the coefficients of $e^{\pm i\delta t}$ on both sides of the equation, then we have 
\begin{eqnarray}
(\frac{\gamma}{2}+i\omega_{m}-i\delta)\delta b_{+}&=&ig_{0}c_{s}^{\ast}\delta c_{+}+ig_{0}c_{s}\delta c_{-}^{\ast},\\
(\frac{\gamma}{2}-i\omega_{m}-i\delta)\delta b_{-}^{\ast}&=&-ig_{0}c_{s}^{\ast}\delta c_{+}-ig_{0}c_{s}\delta c_{-}^{\ast},\\
(2\kappa+i\Delta-i\delta)\delta c_{+}&=&ig_{0}c_{s}(\delta b_{+}+\delta b_{-}^{\ast})+\varepsilon_{p},\\
(2\kappa-i\Delta-i\delta)\delta c_{-}^{\ast}&=&-ig_{0}c_{s}^{\ast}(\delta b_{+}+\delta b_{-}^{\ast}).
\end{eqnarray}
From Eqs. (4)--(7), we can easily obtain
\begin{eqnarray}
\delta c_{+}=\frac{\varepsilon_{p}}{2\kappa-i(\delta-\Delta)+\frac{g_{0}^{2}|c_{s}|^{2}}{Z\times(\frac{\gamma}{2}-i\delta+i\omega_{m})-\frac{g_{0}^{2}|c_{s}|^{2}}{2\kappa-i(\delta+\Delta)}}},
\end{eqnarray}
here, $Z=\frac{i\delta+i\omega_{m}-\frac{\gamma}{2}}{2i\omega_{m}}$.

\section{Optomechanically induced  ultraslow light}

According the input-output relation \cite{Huang2010_041803,Walls}, the quadrature of the  optical components with frequency $\omega_{p}$ in the output field
can be defined as $\varepsilon_{T}=2\kappa\delta c_{+}/\varepsilon_{p}$ \cite{Huang2010_041803}. The real part $\mathrm{Re}[\varepsilon_{T}]$ and imaginary part $\mathrm{Im}[\varepsilon_{T}]$ represent
the absorptive and dispersive behavior of the optomechanical system to the probe field, respectively. Because it is known that the coupling between the cavity and the resonator is strong at the near-resonant frequency, here we consider $\delta\sim\Delta\sim\omega_{m}$.  In the following, we mainly focus on the most studied regime in cavity optomechanics where $\gamma\ll\kappa$, $\omega_{m}$ (then $Z=1+\frac{i\gamma}{4\omega_{m}}\simeq1$) and set $x=\delta-\omega_{m}$. Then, according to Eq. (8), $\varepsilon_{T}$ can be obtained as
\begin{eqnarray}
\varepsilon_{T}=\frac{2\kappa}{2\kappa-ix+\frac{\beta}{\frac{\gamma}{2}-ix+\mathcal{N}}}
\end{eqnarray}
where 
\begin{eqnarray}
\beta&=&\frac{g^{2}_{0}\varepsilon^{2}_{c}}{4\kappa^{2}+\omega^{2}_{m}},\\
\mathcal{N}&=&-\frac{\beta}{2\kappa-2i\omega_{m}}.
\end{eqnarray}
It can be seen from Eq. (10) that $\beta$ depends on the power of the coupling field.
The term $\mathcal{N}$ is the key term and it can strongly affect the properties of the optomechanically induced transparency, and then those of the slow light in the system. Note that if we apply rotating wave approximation to solve the Eq. (3), then the term $\mathcal{N}$ will not exist, see Appendix A for details. Hence the origin of the term $\mathcal{N}$ in Eq. (9) should be effects of non-rotating wave approximation (the explanation of the origin in Ref. \cite{Yan2020pra} is not appropriate).

\begin{figure}[ptb]
	\includegraphics[width=0.45\textwidth]{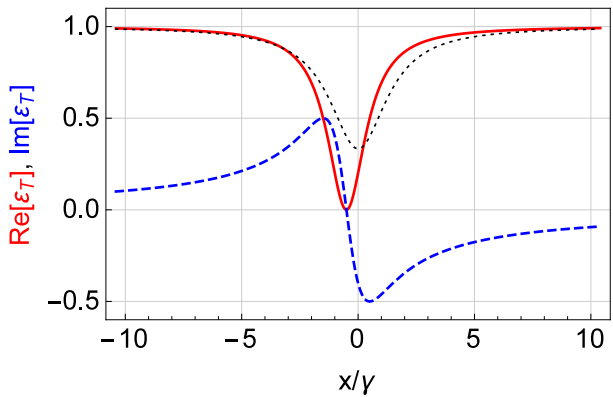}\caption{The real part of $\varepsilon_{T}$ (red-solid) and the  imaginary part of $\varepsilon_{T}$ (blue-dashed) vs. normalized frequency detuning $x/\gamma$ with parameters $\omega_{m}=\kappa=10^{4}\gamma$ according to Eqs. (9) and (13). The real part of $\varepsilon_{T}$ (black-dotted) is plotted with the same parameters but $\mathcal{N}=0$.}%
	\label{Fig2}%
\end{figure}

Since the optomechanically induced slow light is based on the technology of OMIT, we first give the properties of OMIT in the system. How the term $\mathcal{N}$ to affect the OMIT has been studied in detail in our previous work \cite{Yan2020pra}. According to the analysis in Ref. \cite{Yan2020pra}, the location of the pole in the subfraction of Eq. (9) can give the conditions of the ideal OMIT dip. According to the location of the pole, i.e., by 
setting $\frac{\gamma}{2}-ix+\mathcal{N}=0$, we obtain the conditions as
\begin{eqnarray}
x&=&-\frac{\gamma\omega_{m}}{2\kappa},\\
\beta&=&\frac{\gamma(\kappa^{2}+\omega^{2}_{m})}{\kappa}.
\end{eqnarray}
It means that with the driving strength $\beta$ in Eq. (13), the NRWA effect (the term $\mathcal{N}$) just can balance out the mechanical damping rate resulting the ideal OMIT dip.
According to the analysis in Ref. \cite{Yan2020pra} and with the driving strength $\beta$ in Eq. (13), we obtain the width $\Gamma_{\mathrm{OMIT}}$ (full width at half maximum) of transparent window as
\begin{eqnarray}
\Gamma_{\mathrm{OMIT}}=\frac{\gamma(\kappa^{2}+\omega^{2}_{m})}{\kappa^{2}},
\end{eqnarray}
if $\gamma\omega^{2}_{m}\ll4\kappa^{3}$ which is true in most optomechanical systems.
In Fig. 2, we plot $\mathrm{Re}[\varepsilon_{T}]$ (red-solid) and $\mathrm{Im}[\varepsilon_{T}]$ (blue-dashed) vs. normalized frequency detuning $x/\gamma$ with parameters $\omega_{m}=\kappa=10^{4}\gamma$ and $\beta$ according to Eq. (13).
With these parameters, the ideal OMIT dip occurs at detuning $x=-\frac{\gamma}{2}$ and the full width $\Gamma_{\mathrm{OMIT}}=2\gamma$ which
shows an excellent agreement with the numerical result in Fig. 2 (see the red-solid line).
For comparison, we also plot $\mathrm{Re}[\varepsilon_{T}]$ (black-dotted) with the same parameters but $\mathcal{N}=0$ in Fig. 2, from which it can be seen that the ideal OMIT dip cannot occur if the term $\mathcal{N}$ is ignored. 
It can also be seen from Fig. 2 that the steepest dispersion (blue-dashed) appears at transparent window and the slope is negative there.

In Fig. 3, we plot the normalized dispersion curve slope $\gamma\mathrm{K}$ (blue-dashed) vs. normalized frequency detuning $x/\gamma$ with parameters $\kappa=\omega_{m}=10^{4}\gamma$ according to Eqs. (9) and (13). It can be seen from Fig. 3 that the negative maximum value of the dispersion curve slope appears exactly at transparent window ($x=-\frac{\gamma\omega_{m}}{2\kappa}=-\frac{\gamma}{2}$, see the location of the cyan vertical line). According to Eq. (9), the negative maximum value
of the dispersion curve slope can be obtained as
\begin{eqnarray}
\mathrm{K_{max}}=-\frac{2\kappa^{2}}{\gamma(\kappa^{2}+\omega^{2}_{m})}.
\end{eqnarray}
From Eqs. (14) and (15), we have $\mathrm{K_{max}}\times\Gamma_{\mathrm{OMIT}}=-2$, which means that the narrower the width of transparent window is, the steeper the dispersion curve becomes. Next, we will see that the steep dispersion behavior can cause the ultraslow light in the system.  

\begin{figure}[ptb]
	\includegraphics[width=0.46\textwidth]{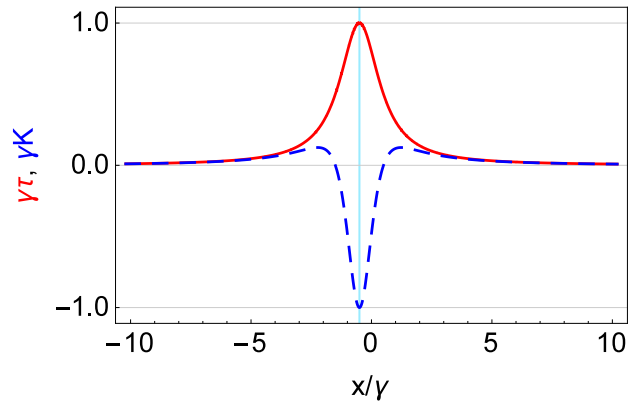}\caption{The normalized time delay $\gamma\tau$ (red-solid) and the  normalized dispersion curve slope $\gamma\mathrm{K}$ (blue-dashed) vs. normalized frequency detuning $x/\gamma$ with parameters $\omega_{m}=\kappa=10^{4}\gamma$ according to Eqs. (9) and (13). The cyan vertical line locates at the transparency window $x=-\frac{\gamma\omega_{m}}{2\kappa}$.}%
	\label{Fig3}%
\end{figure}

The time delays of the transmission and reflection pulses can
be respectively determined by \cite{Weis2010,Safavi-Naeini2011}
\begin{eqnarray}
\tau_{T}=\frac{\partial\mathrm{arg}[\varepsilon_{T}]}{\partial\omega_{p}},\quad 
\tau_{R}=\frac{\partial\mathrm{arg}[\varepsilon_{R}]}{\partial\omega_{p}}.
\end{eqnarray}
Here, $\varepsilon_{R}$ is the reflected component of the probe field, which can be obtained as $\varepsilon_{R}=\varepsilon_{T}-1$ according to input-output relation \cite{Huang2010_041803,Tarhan2013,Walls}. The positive (negative) value of the time delays represents slow (fast) light \cite{Bigelow2003sci} in the system. 
According to Eqs. (9), (13) and (16), we obtain the analytic expressions of time delays $\tau_{T}$ and $\tau_{R}$, and find they are exactly equal ($\tau_{T}=\tau_{R}=\tau$), that is
\begin{eqnarray}
\tau=\frac{8\kappa^2(y^2\kappa+\gamma (\kappa^2+\omega_{m}^2))}{16y^2\kappa^4+ (y\gamma\omega_{m}-2y^2\kappa+2\gamma(\kappa^2 +\omega_{m}^2))^2},
\end{eqnarray}
here $y=x+\frac{\gamma\omega_{m}}{2\kappa}$.
It can be seen from Eq. (17) that there is only slow light effect in the system because the time delay $\tau$ is always positive.

In Fig. 3, we plot the normalized time delay $\gamma\tau$ (red-solid) vs. normalized frequency detuning $x/\gamma$ with parameters $\kappa=\omega_{m}=10^{4}\gamma$ according to Eq. (17). It can be seen from Fig. 3 that the maximum time delay $\tau_{\mathrm{max}}$ occurs exactly at the transparent window $x=-\frac{\gamma\omega_{m}}{2\kappa}=-\frac{\gamma}{2}$. By setting $y=0$ in Eq. (17),
the analytic expression of the maximum time delay $\tau_{\mathrm{max}}$ can be easily obtained as
\begin{eqnarray}
\tau_{\mathrm{max}}=\frac{2\kappa^{2}}{\gamma(\kappa^{2}+\omega^{2}_{m})}.
\end{eqnarray} 
According to Eq. (15) and (18), at the transparent window, we have 
\begin{eqnarray}
\tau_{\mathrm{max}}=-\mathrm{K_{max}}
\end{eqnarray}
which means that the steeper the slope of dispersion curve is, the larger the slow light effect becomes. 

It can be seen from Eq. (18) that the ultraslow light can be achieved by adopting the mechanical oscillator with small enough dissipation rate $\gamma$, especially in unresolved sideband regime. In fact, even with the usual Hz linewidth, the time delay can be largely enhanced. Such as, for $\gamma=0.76$ Hz in Ref. \cite{Thompson2008} and $\kappa=\omega_{m}$, the time delay $\tau_{\mathrm{max}}$ is about one second which is similar to the storage time in Bose-Einstein Condensates \cite{Zhang2009prl} .

Actually, according to Eq. (18) and the mechanical ringdown time $\tau_{m}=2/\gamma$ (in fact, to determine the quality factor of the mechanical resonator, the experimenters perform a mechanical ringdown by suddenly switching off a near-resonant drive and monitoring the amplitude decay. The mechanical ringdown time is the time required for the normalized amplitude to decay form 1 to $e^{-1}$, see Refs. \cite{Ghadimi2018sci,Tsaturyan2017NatN,Thompson2008} for more details), we have
\begin{eqnarray}
\tau_{\mathrm{max}}=\tau_{m}
\end{eqnarray}
in the limit of $\omega_{m}/\kappa\rightarrow0$. Hence, the upper bound of the time delay is exactly the mechanical ringdown time $\tau_{m}$ in the system. Recently, the mechanical ringdown time $\tau_{m}$ can last for several minutes (mHz linewidth) in the experiments, see Refs. \cite{Norte2016prl,Reinhardt2016prx,Ghadimi2018sci,Tsaturyan2017NatN}. Maybe someday in the future, these ultralong time delays
can be used for OMIT-based memories.

\section{Optomechanically induced transmission}

Now we study the interesting phenomenon of perfect optomechanically induced transmission which can be realized with the term $\mathcal{N}$. If the reflected component $\varepsilon_{R}=0$ and the transmitted component $\varepsilon_{T}=1$, the perfect optomechanically induced transmission of the probe signal will occur. According to Eq. (9), this can be achieved when 
\begin{eqnarray}
-ix+\frac{\beta}{\frac{\gamma}{2}-ix+\mathcal{N}}=0.
\end{eqnarray}
From Eq. (21), we find the perfect optomechanically induced transmission can occur only when
\begin{eqnarray}
\beta&=&\frac{\gamma(\kappa^{2}+\omega_{m}^{2})}{\kappa},\\
x_{\pm}&=&\frac{-\gamma\omega_{m}\pm\sqrt{\gamma(16\kappa^{3}+16\kappa\omega_{m}^{2}+\gamma\omega_{m}^{2})}}{4\kappa}.
\end{eqnarray} 
Here $x_{\pm}$ can be simplified to $\pm\sqrt{\gamma(\kappa^{2}+\omega_{m}^{2})/\kappa}$
due to $\gamma\ll\kappa$, $\omega_{m}$. 
The driving strength $\beta$ in Eq. (22) is exactly equal to that in Eq. (13), which means that the optomechanically induced transparency and the optomechanically induced transmission happen with the same driving strength, just at different frequencies of the probe signal. Note that if the term $\mathcal{N}$ is ignored, the perfect optomechanically induced transmission cannot occur in the system.

In Fig. 4, we plot the reflection spectrum $R=|\varepsilon_{T}-1|^2$ (blue-dashed) and transmission
spectrum $T=|\varepsilon_{T}|^2$ (red-solid) of the probe field vs. normalized frequency detuning $x/\gamma$ with $\beta$ according to Eq. (13) and the parameters $\omega_{m}=10\kappa=10^{4}\gamma$. It can be seen from Fig. 4 that the probe signal can be perfectly reflected ($R=1$) at the transparency window $x=-\frac{\gamma\omega_{m}}{2\kappa}$, while it can be perfectly transmitted ($T=1$) at the transmission points $x_{\pm}$ in Eq. (23). With these parameters, $x_{\pm}\simeq\pm\sqrt{\gamma(\kappa^{2}+\omega_{m}^{2})/\kappa}\simeq\pm317.8\gamma$ which are consistent with the numerical results in Fig. 4, see the location of the cyan vertical lines plotted according to Eq. (23).

\begin{figure}[ptb]
	\includegraphics[width=0.45\textwidth]{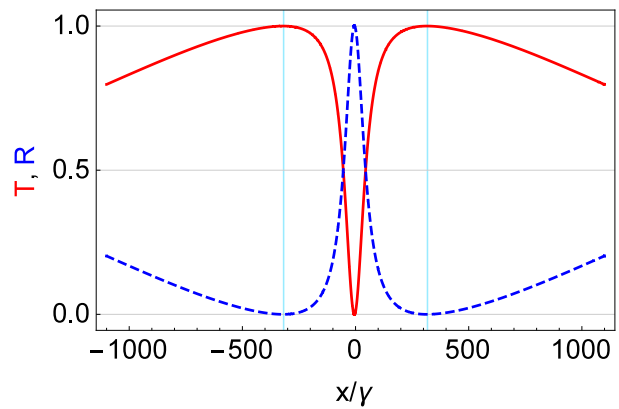}\caption{The reflection spectrum $R$ (blue-dashed) and transmission spectrum $T$ (red-solid) vs. normalized frequency detuning $x/\gamma$ with $\beta$ according to Eq. (13) and parameters $\omega_{m}=10\kappa=10^{4}\gamma$. The cyan vertical lines locate at $x_{\pm}$ according to Eq. (23).}%
	\label{Fig4}%
\end{figure}

According to Eq. (9), we can obtain the slopes of $\mathrm{Im}[\varepsilon_{T}]$ at $x_{\pm}$ as
\begin{eqnarray}
\mathrm{K}_{\pm}=\frac{1}{\kappa\pm\sqrt{\frac{\gamma\kappa^{2}\omega_{m}^{2}}{16\kappa^{3}+\gamma\omega_{m}^{2}+16\kappa\omega_{m}^{2}}}},
\end{eqnarray}
which can be simplified to $1/\kappa$
due to $\gamma\ll\kappa,$ $\omega_{m}.$ From Eq. (17), the 
time delays $\tau_{\pm}$ at the transmission point $x_{\pm}$ can be obtained as
\begin{eqnarray}
\tau_{\pm}=\mathrm{K}_{\pm}\simeq\frac{1}{\kappa}.
\end{eqnarray}
It means that compared with the mechanical ringdown time $\tau_{m}$, there is almost no time delay at the transmission points $x_{\pm}$ due to $\gamma\ll\kappa$.

\section{Optomechanically induced absorption}

Now we study the interesting phenomenon of perfect optomechanically induced absorption (OMIA) which can also be strongly affected by the term $\mathcal{N}$. For this purpose,
two probe fields with amplitudes $\varepsilon_{l},$ $\varepsilon_{r}$ and identical frequency $\omega_{p}$ are respectively injected to the system from two sides, see Fig. 5. The two output fields are denoted by $\varepsilon_{outL}$ and $\varepsilon_{outR}$ and the other parameters are denoted as the same as above. When the perfect OMIA occurs, the two probe fields are completely absorbed by the system so that there is no light field with frequency $\omega_{p}$ in the output fields $\varepsilon_{outL}$ and $\varepsilon_{outR}$.

\begin{figure}[ptb]
	\includegraphics[width=0.41\textwidth]{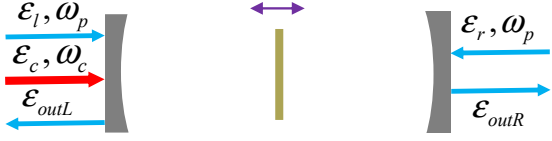}\caption{Two probe fields with amplitudes $\varepsilon_{l},$ $\varepsilon_{r}$ and identical frequency $\omega_{p}$ are injected to the system from two sides. The two output fields are denoted by $\varepsilon_{outL}$ and $\varepsilon_{outR}$ and the other parameters are the same in Fig. 1.}%
	\label{Fig5}%
\end{figure}

In this case, with the similar calculation as above, we have
\begin{eqnarray}
\delta c_{+}=\frac{\varepsilon_{l}+\varepsilon_{r}}{2\kappa-ix+\frac{\beta}{\frac{\gamma}{2}-ix+\mathcal{N}}}.
\end{eqnarray}
To study the perfect OMIA, we first calculate the output fields of the two sides of the cavity at frequency $\omega_{p}$, which can be derived by the input-output relations \cite{Huang2010_041803,Agarwal2014njp}
\begin{eqnarray}
\varepsilon_{outL}+\varepsilon_{l}e^{-i\delta t}=2\kappa\delta c,\notag\\
\varepsilon_{outR}+\varepsilon_{r}e^{-i\delta t}=2\kappa\delta c.
\end{eqnarray}
Similarly, we can write the output fields as
\begin{eqnarray}
\varepsilon_{outL}=\varepsilon_{outL+}e^{-i\delta t}+\varepsilon_{outL-}e^{i\delta t},\notag\\
\varepsilon_{outR}=\varepsilon_{outR+}e^{-i\delta t}+\varepsilon_{outR-}e^{i\delta t},
\end{eqnarray}
where $\varepsilon_{outL+}$ and $\varepsilon_{outR+}$ are oscillating at frequency $\omega_{p}$ in the output fields. Thus, when $\varepsilon_{outL+}=\varepsilon_{outR+}=0$, the perfect OMIA will occur. From Eqs. (27) and (28), we have
\begin{eqnarray}
\varepsilon_{outL+}=2\kappa\delta c_{+}-\varepsilon_{l},\notag\\
\varepsilon_{outR+}=2\kappa\delta c_{+}-\varepsilon_{r}.
\end{eqnarray}

According to Eq. (26) and (29) and due to $\gamma\ll\kappa$, $\omega_{m}$, the conditions of the perfect OMIA can be obtained as
\begin{eqnarray}
\varepsilon_{l}&=&\varepsilon_{r},\\
\beta&=&\frac{\gamma\kappa(\kappa^{2}+\omega_{m}^{2})}{2\kappa^{2}+\omega_{m}^{2}},\\
x&=&-\frac{\gamma\kappa\omega_{m}}{4\kappa^{2}+2\omega_{m}^{2}},
\end{eqnarray} 
which can be verified in Fig. 6 where we plot the normalized output probe field $|\varepsilon_{outR+}/\varepsilon_{r}|^{2}$ ($|\varepsilon_{outL+}/\varepsilon_{l}|^{2}$) vs. normalized frequency detuning $x/\gamma$ according to Eqs. (29)--(31) with resolved sideband parameter $\omega_{m}=10\kappa=10^{4}\gamma$ [red-solid ($\mathcal{N}\neq0$) and black-dash-dotted ($\mathcal{N}=0$)] and unresolved sideband parameter $\omega_{m}=\kappa/2=10^{4}\gamma$ [blue-dashed ($\mathcal{N}\neq0$) and green-dotted ($\mathcal{N}=0$)]. For comparison, we also plot the same spectrum with the same parameters but $\mathcal{N}=0$, see the black-dash-dotted line and the green-dotted line in Fig. 6. Through comparison, we find that the term $\mathcal{N}$ can strongly affect the perfect OMIA in unresolved sideband regime while it has almost no effect in resolved sideband regime ($\kappa\ll\omega_{m}$).

\begin{figure}[ptb]
	\includegraphics[width=0.45\textwidth]{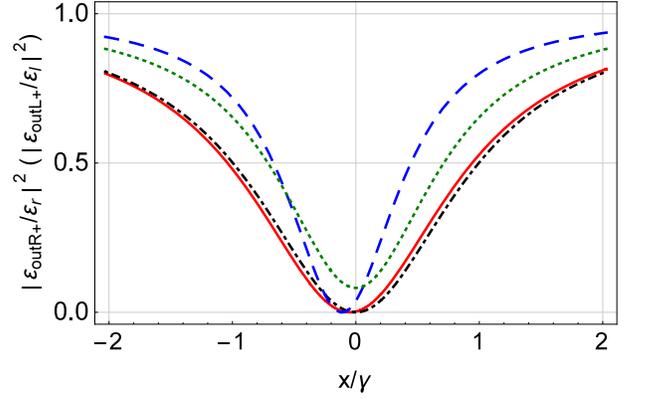}\caption{The normalized output probe field $|\varepsilon_{outR+}/\varepsilon_{r}|^{2}$ ($|\varepsilon_{outL+}/\varepsilon_{l}|^{2}$) vs. normalized frequency detuning $x/\gamma$ according to Eqs. (29)--(31) with resolved sideband parameter $\omega_{m}=10\kappa=10^{4}\gamma$ [red-solid ($\mathcal{N}\neq0$) and black-dash-dotted ($\mathcal{N}=0$)] and unresolved sideband parameter $\omega_{m}=\kappa/2=10^{4}\gamma$ [blue-dashed ($\mathcal{N}\neq0$) and green-dotted ($\mathcal{N}=0$)].}%
	\label{Fig6}%
\end{figure}

Finally, we study the perfect OMIA in resolved sideband regime. As mentioned above, in this case, the term $\mathcal{N}$ can be ignored. Actually, this issue has been studied in Ref. \cite{Agarwal2014njp} where the authors give a condition to realize the perfect OMIA, i.e., $\varepsilon_{l}=\varepsilon_{r},$ $\gamma=4\kappa$ and $x=\pm\sqrt{\beta-4\kappa^{2}}$, which means that the driving strength $\beta$ must be equal or greater than $4\kappa^{2}$ and the mechanical damping rate $\gamma$ can take only one value $4\kappa$ (actually, this is a huge damping rate to mechanical oscillator in cavity optomechanics). Here, we give another simple but interesting condition to realize the perfect OMIA, that is
\begin{eqnarray}
\varepsilon_{l}&=&\varepsilon_{r},\notag\\
\beta&=&\kappa\gamma,\\
x&=&0,\notag
\end{eqnarray}
which can be easily obtained according to Eqs. (26) and (29) with $\mathcal{N}=0$.
It can be seen from Eq. (33) that the mechanical damping rate $\gamma$ can take any value to realize the perfect OMIA as long as the conditions in Eq. (33) is satisfied.
In Fig. 7, we plot the normalized output probe field $|\varepsilon_{outR+}/\varepsilon_{r}|^{2}$ ($|\varepsilon_{outL+}/\varepsilon_{l}|^{2}$) vs. normalized frequency detuning $x/\kappa$ under the conditions in Eq. (33) and with the resolved sideband parameter $\omega_{m}=10\kappa$ and $\gamma=\kappa/10$ (red-solid), $\gamma=\kappa$ (black-dotted) and $\gamma=2\kappa$ (blue-dashed).
It can be seen from Fig. 7 that the perfect OMIA can still be achieved even with very weak driving strength ($\beta\ll\kappa^{2}$) [see the red-solid line ($\beta=\kappa^{2}/10$)].

\begin{figure}[ptb]
	\includegraphics[width=0.45\textwidth]{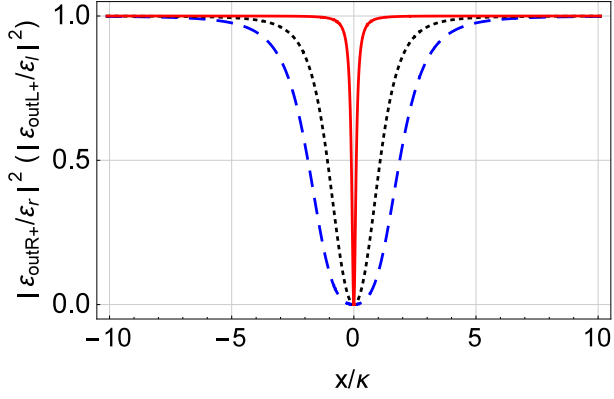}\caption{The normalized output probe field $|\varepsilon_{outR+}/\varepsilon_{r}|^{2}$ ($|\varepsilon_{outL+}/\varepsilon_{l}|^{2}$) vs. normalized frequency detuning $x/\kappa$ under the conditions in Eq. (33) and with the resolved sideband parameter $\omega_{m}=10\kappa$ and $\gamma=\kappa/10$ (red-solid), $\gamma=\kappa$ (black-dotted) and $\gamma=2\kappa$ (blue-dashed).}%
	\label{Fig7}%
\end{figure}

\section{Conclusion}

In summary, we have theoretically studied the optomechanically induced ultraslow light, perfect transmission and absorption in a membrane-in-the-middle optomechanical system with an effect of non-rotating wave approximation (NRWA) that was ignored in previous works. From the theoretical results, we can draw some important conclusions: (1) With the NRWA effect, the ultraslow light can be easily achieved with a small mechanical damping rate at transparency window, especially in unresolved sideband regime (for instance, the time delay can be enhanced to about one second with Hz linewidth); (2) there exists an upper bound of the time delay, and it is exactly the mechanical ringdown time which can last for several minutes; (3) after considering the NRWA effect, we obtain, respectively, the conditions under which the perfect optomechanically induced transmission and absorption can be achieved in the system. Furthermore, the methods are also applicable to other optomechanical systems, such as superconducting circuits \cite{Regal2008,Teufel2011_471}, multimode optomechanical systems \cite{Yan2014,Qu2013pra,Yan2019OE}, and spinning optomechanical systems \cite{Lu2017}. We believe these results can be used to control optical transmission in quantum information processing.

\appendix{}

\section{Derivation of $\varepsilon_{T}$ with rotating wave approximation}

To find out the origin of the term $\mathcal{N}$, we apply rotating wave approximation to solve Eq. (3). To this end, we introduce the substitutions $\delta b\rightarrow\delta\tilde{b}e^{-i\omega_{m}t}$, $\delta c\rightarrow\delta\tilde{c}e^{-i\Delta t}$, and then Eq. (3) becomes
\begin{eqnarray}
\delta\dot{\tilde{b}}=&-&\frac{\gamma}{2}\delta \tilde{b}+ig_{0}(c_{s}^{\ast}\delta \tilde{c}e^{-i(\Delta-\omega_{m})t}+c_{s}\delta \tilde{c}^{\ast}e^{i(\Delta+\omega_{m})t}),\notag\\
\delta\dot{\tilde{c}}=&-&2\kappa\delta \tilde{c}+ig_{0}c_{s}(\delta \tilde{b}e^{i(\Delta-\omega_{m})t}+\delta \tilde{b}^{\ast}e^{i(\Delta+\omega_{m})t})\notag\\
&+&\varepsilon_{p}e^{-i(\delta-\Delta)t}.
\end{eqnarray}
If the cavity is driven by a coupling field at the
mechanical red sideband $\Delta\simeq\omega_{m}$ and we apply rotating wave approximation, i.e., neglecting the counter-rotating terms (non-resonant contributions), then Eq. (A1) becomes
\begin{eqnarray}
\delta\dot{\tilde{b}}&=&-\frac{\gamma}{2}\delta \tilde{b}+ig_{0}c_{s}^{\ast}\delta \tilde{c},\\
\delta\dot{\tilde{c}}&=&-2\kappa\delta \tilde{c}+ig_{0}c_{s}\delta \tilde{b}+\varepsilon_{p}e^{-i(\delta-\omega_{m})t}.\notag
\end{eqnarray}
Now we introduce the reverse substitutions $\delta \tilde{b}\rightarrow\delta be^{i\omega_{m}t}$, $\delta \tilde{c}\rightarrow\delta ce^{i\omega_{m} t}$, and Eq. (A2) becomes
\begin{eqnarray}
\delta\dot{b}&=&-(\frac{\gamma}{2}+i\omega_{m})\delta b+ig_{0}c_{s}^{\ast}\delta c,\\
\delta\dot{c}&=&-(2\kappa+i\omega_{m})\delta c+ig_{0}c_{s}\delta b+\varepsilon_{p}e^{-i\delta t}.\notag
\end{eqnarray} 
Still following the assumption $\delta b=\delta b_{+}e^{-i\delta t}+\delta b_{-}e^{i\delta t}$ and $\delta c=\delta c_{+}e^{-i\delta t}+\delta c_{-}e^{i\delta t}$, and substituting them into Eq. (A3), then, we can obtain the expression of $\delta c_{+}$ and according to $\varepsilon_{T}=2\kappa\delta c_{+}/\varepsilon_{p}$, $\varepsilon_{T}$ can be obtained as
\begin{eqnarray}
\varepsilon_{T}=\frac{2\kappa}{2\kappa-ix+\frac{\beta}{\frac{\gamma}{2}-ix}},
\end{eqnarray}
here $x=\delta-\omega_{m}$. It can be clearly seen from Eq. (A4) that the term $\mathcal{N}$ in Eq. (9) does not appear here. Hence the origin of the term $\mathcal{N}$ in Eq. (9) should be the effect of non-rotating wave approximation.

\end{document}